\documentclass[aps,pre,reprint,superscriptaddress,showpacs]{revtex4-1}
\usepackage{amsmath}

\usepackage{url}
\usepackage{slashbox}
\usepackage{amsfonts,amssymb,amsthm,graphicx,color,epstopdf,tipa}
\usepackage{booktabs}
\usepackage[table]{xcolor}
\PassOptionsToPackage{hyphens}{url}
\usepackage{hyperref}
\newcommand{\PP}{ \mathbf{Pr} }

\newcommand{\RR}{ \mathbb{R} }

\newcommand{\rar}{\rightarrow}
\newcommand{\bv}[1]{\boldsymbol{\mathbf{#1}}}

\newcommand{\la}{\langle}
\newcommand{\ra}{\rangle}

\begin{document}

%\preprint{}

\title{Statistical Physics of Language Maps in the USA}


\author{J. Burridge}
\email[]{james.burridge@port.ac.uk}

%\thanks{}
\affiliation{School of Mathematics and Physics, University of Portsmouth, UK}

\author{B. Vaux}

\affiliation{Faculty of Medieval and Modern Languages, University of Cambridge, UK}

\author{M. Gnacik}

\author{Y. Grudeva}

\affiliation{School of Mathematics and Physics, University of Portsmouth, UK}


\date{\today}

\begin{abstract}
Spatial linguistic surveys often reveal well defined geographical zones where certain linguistic forms are dominant over their alternatives. It has been suggested that these patterns may be understood by analogy with coarsening in models of two dimensional physical systems. Here we investigate this connection by comparing data from the \textit{Cambridge Online Survey of World Englishes} to the behaviour of a generalised zero temperature Potts model with long range interactions. The relative displacements of linguistically similar population centres reveals enhanced east-west affinity. Cluster analysis reveals three distinct linguistic zones.  We find that when the interaction kernel is made anisotropic by stretching along the east-west axis, the model can reproduce the three linguistic zones for all interaction parameters tested. The model results are consistent with a view held by some linguists that, in the USA, language use is, or has been, exchanged or transmitted to a greater extent along the east-west axis than the north-south.
\end{abstract}

%\keywords{}

\maketitle

\section{Introduction}

All people display linguistic idiosyncrasies \cite{cha98}. These might be different words for the same action or object, syntactical differences, or systematic variations in pronunciation. A speaker's geographical origins can often be inferred from their use of language, because people from the same region typically have many linguistic features in common. For example, native English speakers from western Canada typically call a multistory car park a “parkade”, athletic shoes worn as casual footwear “runners”, and small houses in the countryside for weekend retreats during the summer months “cabins” \cite{bob10}. A collection of particularly consistent and distinctive pronunciations may be called an \textit{accent}, or if vocabulary and grammar are also distinctive, a \textit{dialect}  \cite{cha98}.  The earliest known study of geographical language variation was carried out in 1876 by Georg Wenker, who asked 50,000 schoolmasters from locations across Germany to transcribe a list of sentences into the local dialect \cite{cha92}. Modern computers and the creation of the internet have dramatically improved data collection and analysis \cite{ner03,ner10,ner11,vau17,wie15,hee04,wie11,hee01,gri11}, and social media has provided a new source of linguistic data \cite{gri16}.  Modelling linguistic evolution has also emerged as a sub-field of statistical physics where ideas and techniques employed to relate the macroscopic behaviour of physical systems to their microscopic components have been applied \cite{bax06,ger14,pet12,Cas09,bly07,bur16,bur17,bur18,bur18_2}. However, there is a need to develop mathematical models which provide a scientific understanding of how human-level processes \cite{sta13} give rise to the observed geographical distributions and language dynamics. It has recently been proposed by the lead author \cite{bur17,bur18} that geographical boundaries between linguistic features are analogous to domain walls in physical systems \cite{bra94,bur16,kra10}, causing them to straighten over time, and to be repelled by population centres. These effects lead to significant predictability in the geographical distribution of language use. In this paper, we explore this hypothesis by comparing the behaviour of a simple Potts-type model of language evolution, to a large modern dialect survey of the United States.

\section{Survey Data} 

The Cambridge Online Survey of World Englishes (COSWE) \cite{vau17}, initiated in 2007, consists of geographically located responses to thirty-one different questions. For example, question five is: 

\vspace{0.2cm}
\noindent
\textit{What word(s) do you use in casual speech to address a group of two or more people? } 

\vspace{0.2cm}

\begin{figure}
	\includegraphics[width=\columnwidth]{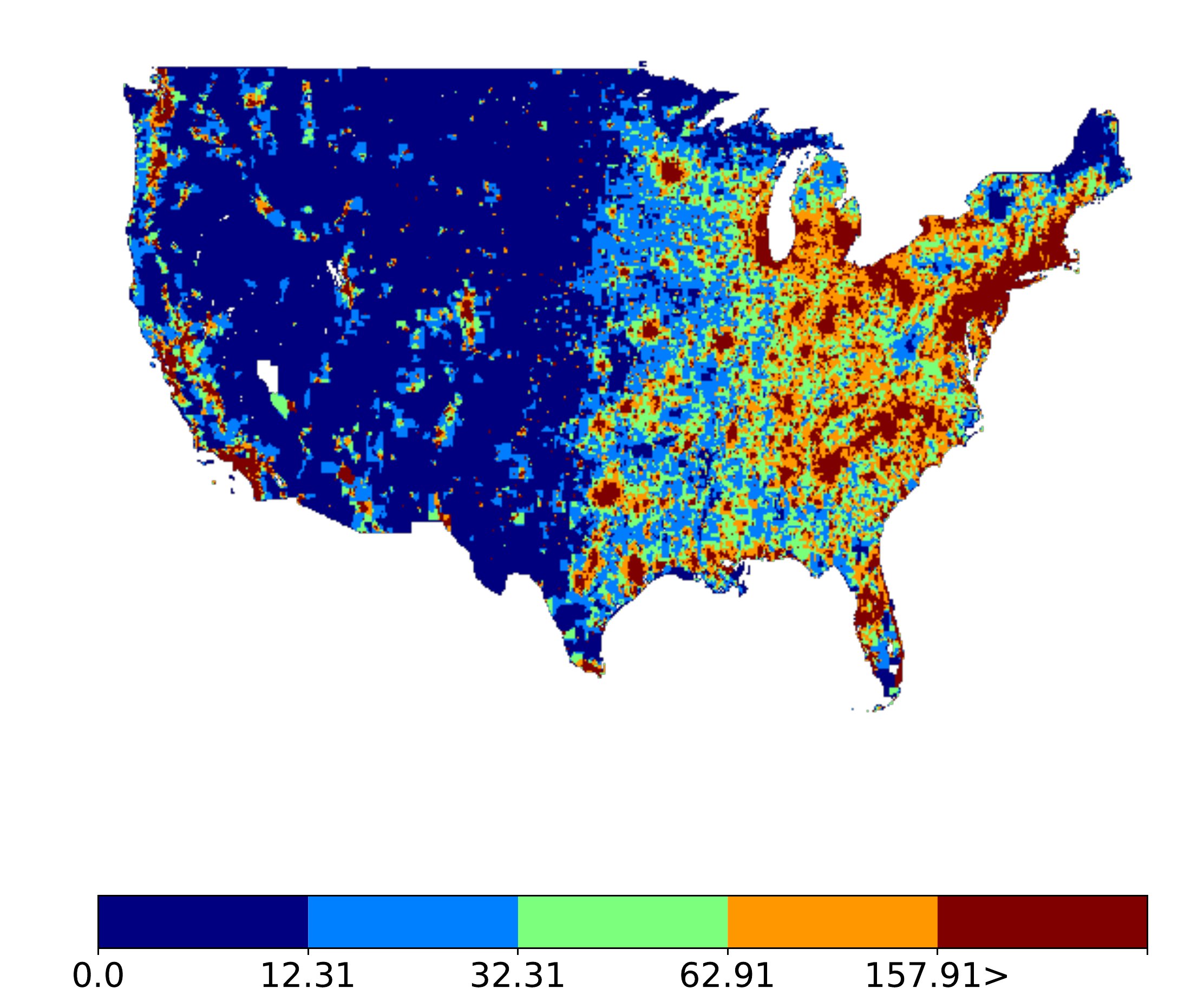}
	\caption{\label{fig:pop_den} Population per square mile for the USA.     }
\end{figure}

\noindent
The most popular answers to this particular question are \textit{you guys} (35\%) and \textit{y'all} (15\%).  The survey currently contains $8.28 \times 10^4$ responses world-wide, with approximately $5.8 \times 10^4$ located in the eastern half of the United States. This region of high population density, stretching from the desert states, up to the Atlantic Coast, has a wide variety of local linguistic terminology and is the focus of our study. The westernmost cities in our study are San Antonio in the South (Texas) and Fargo in the North (Minnesota / North Dakota border). The land to the west of our study area is very sparsely populated ($<4$ people per km\textsuperscript{2}) and we approximate it as empty in our model: we treat our study area as a closed system in a linguistic sense. Further justification for this approach is  provided by the USA population density map in Figure \ref{fig:pop_den}.  From this we see that our study area forms a self contained region of high population density bordered by desert, water (The Great Lakes, the Atlantic and Gulf of Mexico) and country boundaries (Canada and Mexico).

\begin{figure}
	\includegraphics[width=\columnwidth]{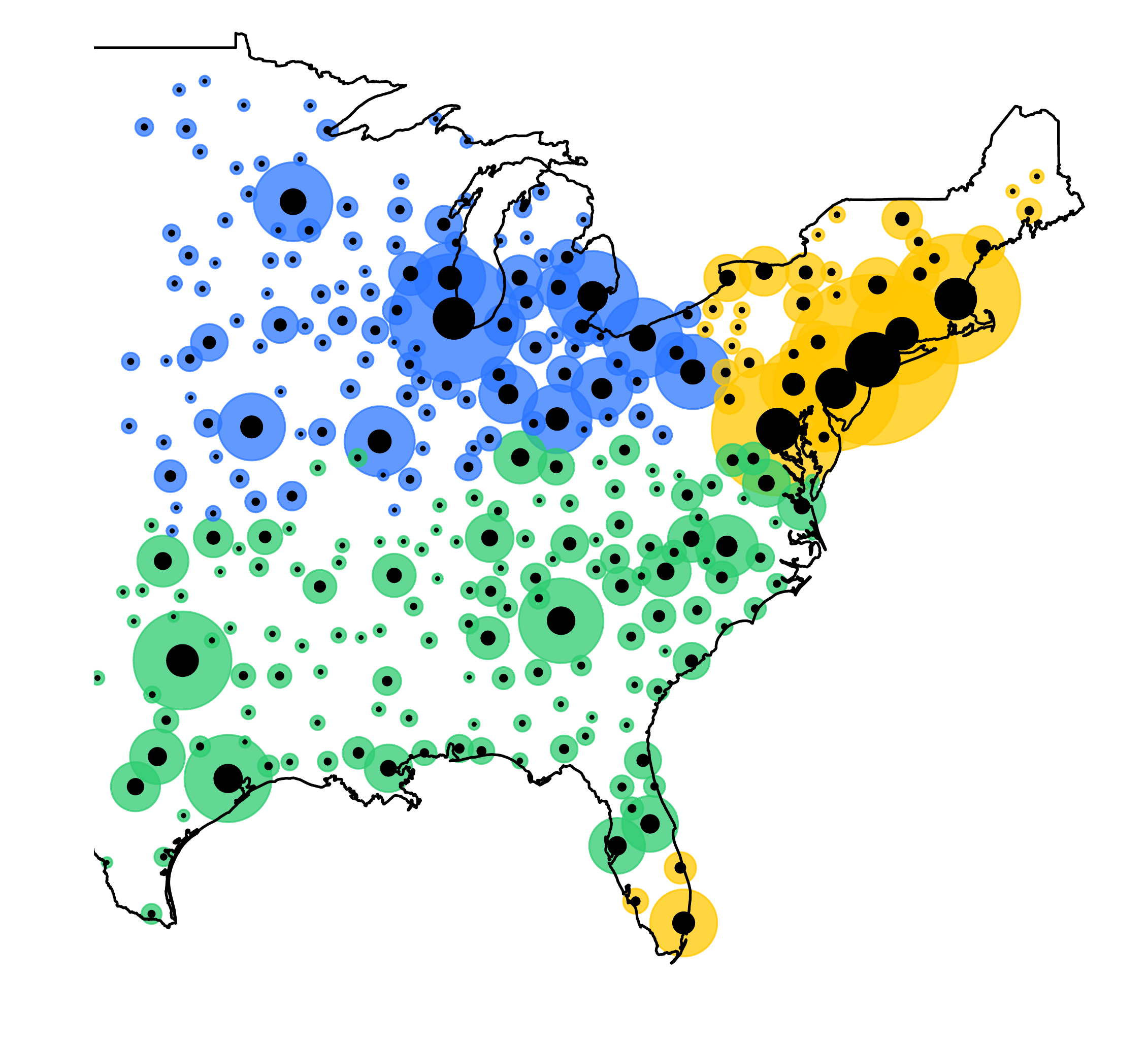}
	\caption{\label{fig:all_data} Our set of 300 population nodes, coloured according to linguistic cluster membership as determined by $K$-means analysis of aggregated survey data.    }
\end{figure}

Maps of the raw survey data \cite{vau17} reveal a patchwork of geographical regions with distinctive language use. However, within any given region linguistic choices are never uniform across all respondents. For example, in Philadelphia, the dominant local term for a submarine sandwich is ``hoagie'', but it is not universally adopted (57\% of people use it in Philadelphia, compared to 8\% in the eastern USA as a whole). In order to characterise local average linguistic behaviour, we begin by clustering the locations of survey respondents using the Mean-Shift algorithm \cite{fuk75}, which locates the peaks of the kernel density estimate of the population distribution. Using a bandwidth of 50 km generates clusters with centroids corresponding to the locations of all significant population settlements within the eastern United States. In regions of low population density which are significantly more than 50 km away from any major settlement we find a large number of small, evenly spaced clusters, each containing only a handful of survey responses ($\ll 20$).  To ensure that each cluster has sufficient data to provide a reliable linguistic sample, we repeatedly join the smallest cluster to its nearest neighbour. We set this minimum sample size to be 20, which is achieved by repeating our joining process until we have $N=300$ nodes (Figure \ref{fig:all_data}). At each node $i$ we define an average frequency vector $\bv{f}^Q_i=(f^Q_{i1},f^Q_{i2}, \ldots f^Q_{i,R_Q})$ for each question, where $f^Q_{ik}$ is the relative frequency of the $k$th response to the $Q$th question, and $R_Q$ is the number of different responses for question $Q$. Each node occupies a point in the \textit{linguistic space} for each question, the probability simplex
\begin{equation}
C_{iQ} = \left\{ \bv{f}^Q_i \in \RR^{R_Q} | f^Q_{i1} + f^Q_{i2} + \ldots + f^Q_{i,R_Q} = 1, f^Q_{ij} \geq 0 \right\} 
\end{equation}
which can be of high dimension -- the COSWE survey database \cite{vau17} for example contains more than 800 distinct families of lexical responses to Question 8, ``What do you call the gooey or dry matter that collects in the corners of your eyes, especially while you are sleeping?'', with the most common being (eye) boogers, sleep, (eye) gunk, and (eye) crusties.   

\begin{figure}
	\includegraphics[width=\columnwidth]{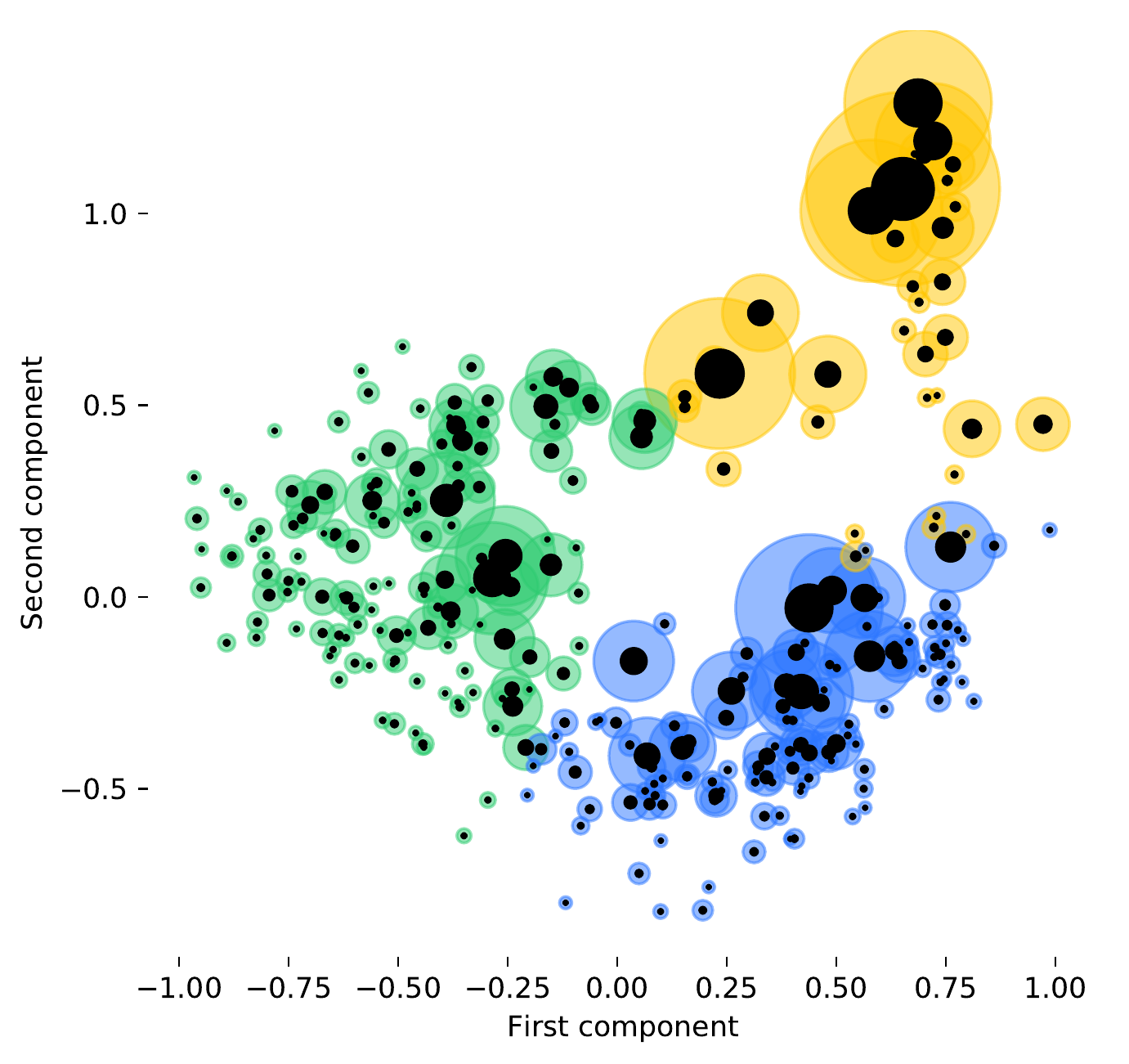}
	\caption{\label{fig:pca} Projection of the node-average combined frequency vectors for all survey questions onto the first two principle component loading vectors. This projection captures $ \approx 50\%$ of the variation. Dot sizes are proportional to node sample sizes. The loading vectors are orthogonal unit vectors, $\bv{v}_1, \bv{v}_2 \ldots$ in linguistic space constructed iteratively, beginning with $\bv{v}_1$. Vector $\bv{v}_k$ is chosen to maximize the variance of $ \bv{v}_k \cdot \bv{f}^i$ over all nodes subject to the condition that $\bv{v}_k \cdot \bv{v}_j=0$ for all $j<k$. Colours give the result of clustering the data into three clusters using $K$-means.   }
\end{figure}

In order to visualize the distribution of population nodes in linguistic space we perform a principle components analysis \cite{has09} to find a low dimensional representation of the frequency data which captures as much of its variation as possible. The result of this analysis for the combined frequencies for all questions $\bv{f}_i = (\bv{f}_i^1, \bv{f}_i^2, \ldots , \bv{f}_i^{n_Q} )$ at each node is shown in Fig. \ref{fig:pca}.
The position of each black dot in figure \ref{fig:pca} represents the linguistic state of a population node, and we note that nodes exhibit a significant degree of clustering. To analyse these clusters we define the linguistic distance between nodes 
\begin{align}
l_{ij} &= |\bv{f}_i-\bv{f}_j| \\
&= \sqrt{\sum_{Q=1}^{n_Q} \left|\bv{f}_i^Q - \bv{f}_j^Q\right|^2},
\end{align}
allowing us to use the $K$-means method \cite{has09} to divide the data into $q \geq 1$ linguistic clusters. We determine the optimal $q$ by maximizing the average silhouette score, $U$, \cite{rou87} over all nodes $i$ where
\begin{align}
U_i &= \frac{b_i-a_i}{\max\{a_i, b_i\} } \in [-1,1] \\
U & = \frac{1}{N} \sum_{i=1}^N U_i
\end{align}
with $a_i$ the average distance between $i$ and nodes in the same cluster, and $b_i$ the average distance to nodes in different clusters. For the aggregated data shown in Figure \ref{fig:pca} we find $q=3$, and nodes are coloured according to their $K$-means cluster label. For comparison, the $q$ values and their scores, $U$, were $(q=2,s=0.21),(3,0.26),(4,0.19),\ldots$ with lower scores for larger $q$ values. Visual inspection of the clusters in Figure \ref{fig:pca} also suggests that $q=3$ is the appropriate choice. 

\begin{figure}
	\includegraphics[width=\columnwidth]{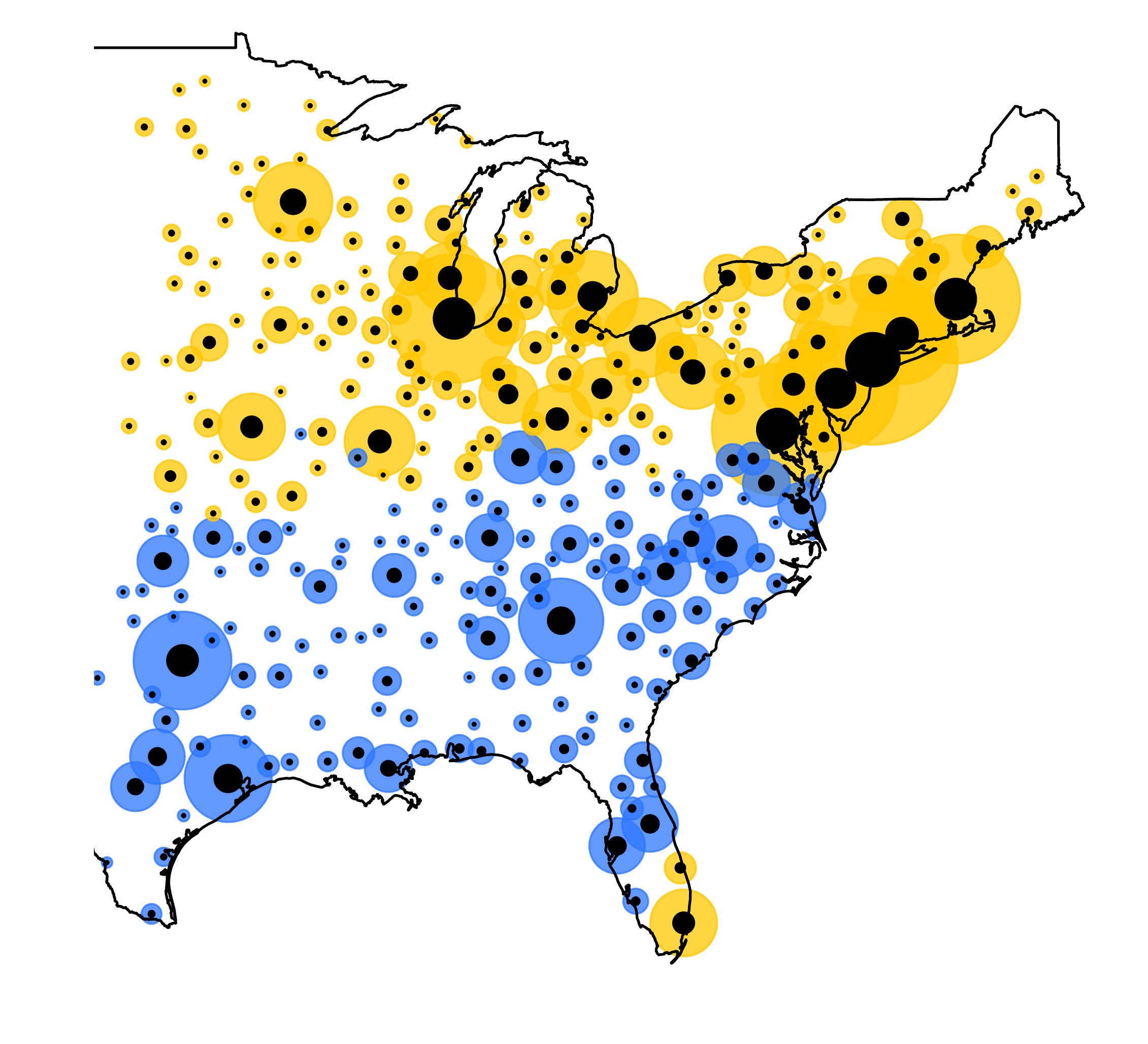}
	\caption{\label{fig:yall} Black dots show locations and sizes of population nodes determined by Mean-Shift clustering (bandwidth 50 kilometres) and subsequent merging of the smallest clusters.  Colours show the results of $K$-means clustering of the responses to question five into $q=2$ clusters. Blue: mostly ``y'all'', Yellow: mostly ``You guys''.       }
\end{figure}

\begin{figure}
	\includegraphics[width=\columnwidth]{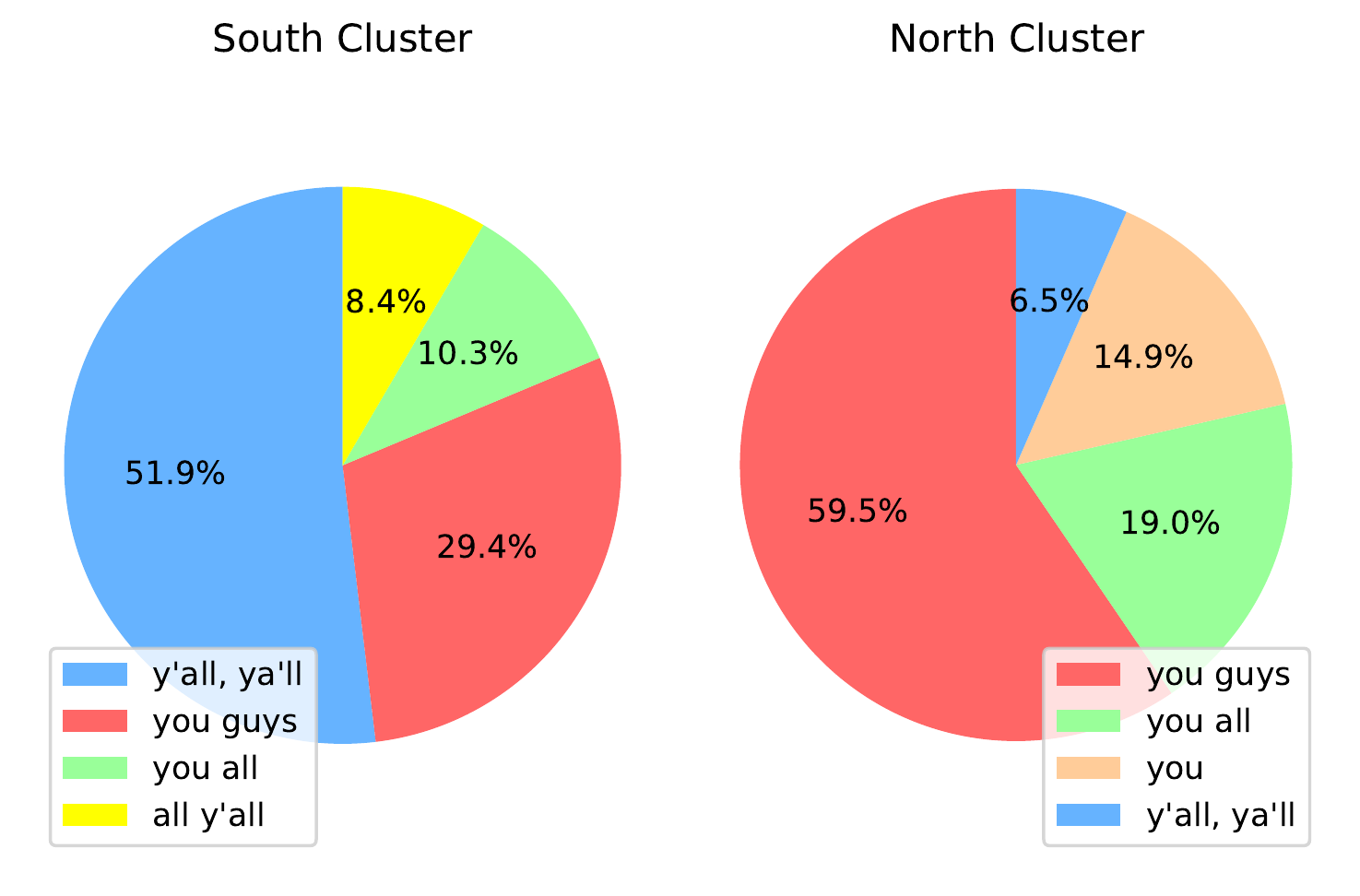}
	\caption{\label{fig:pie} Proportions of different survey responses to question 5 for each of the two clusters shown in Figure \ref{fig:yall}. }
\end{figure}

Node colours in Fig. \ref{fig:all_data} show how the results of our linguistic cluster analysis appear in geographical space. This map demonstrates that geographical proximity is a powerful predictor of linguistic similarity. The linguistic clustering process took no account of geographical location, and yet the resulting clusters divide the spatial domain into distinct regions. Similar divisions appear on the level of individual questions. For example, in Figure \ref{fig:yall} we have performed a linguistic clustering of the responses to question 5, and we see a sharp transition between the Southern states, where groups of people are typically addressed using the expression ``y'all'', and Northern states, where the term is more typically ``you guys''. 
The breakdown of survey results in the two clusters is given in Figure \ref{fig:pie}.  
In the linguistic context, domain walls of this type are known as \textit{isoglosses} \cite{blo33,cha98}, and have been mapped and studied for over a century. Domain walls are also ubiquitous in atomic level ordering processes, and this connection between physics and linguistics was recently reported and explored in detail in \cite{bur17,bur18}. The work we present here is the first quantitative test of this idea against a large linguistic data set. 

\begin{figure}
	\includegraphics[width=\columnwidth]{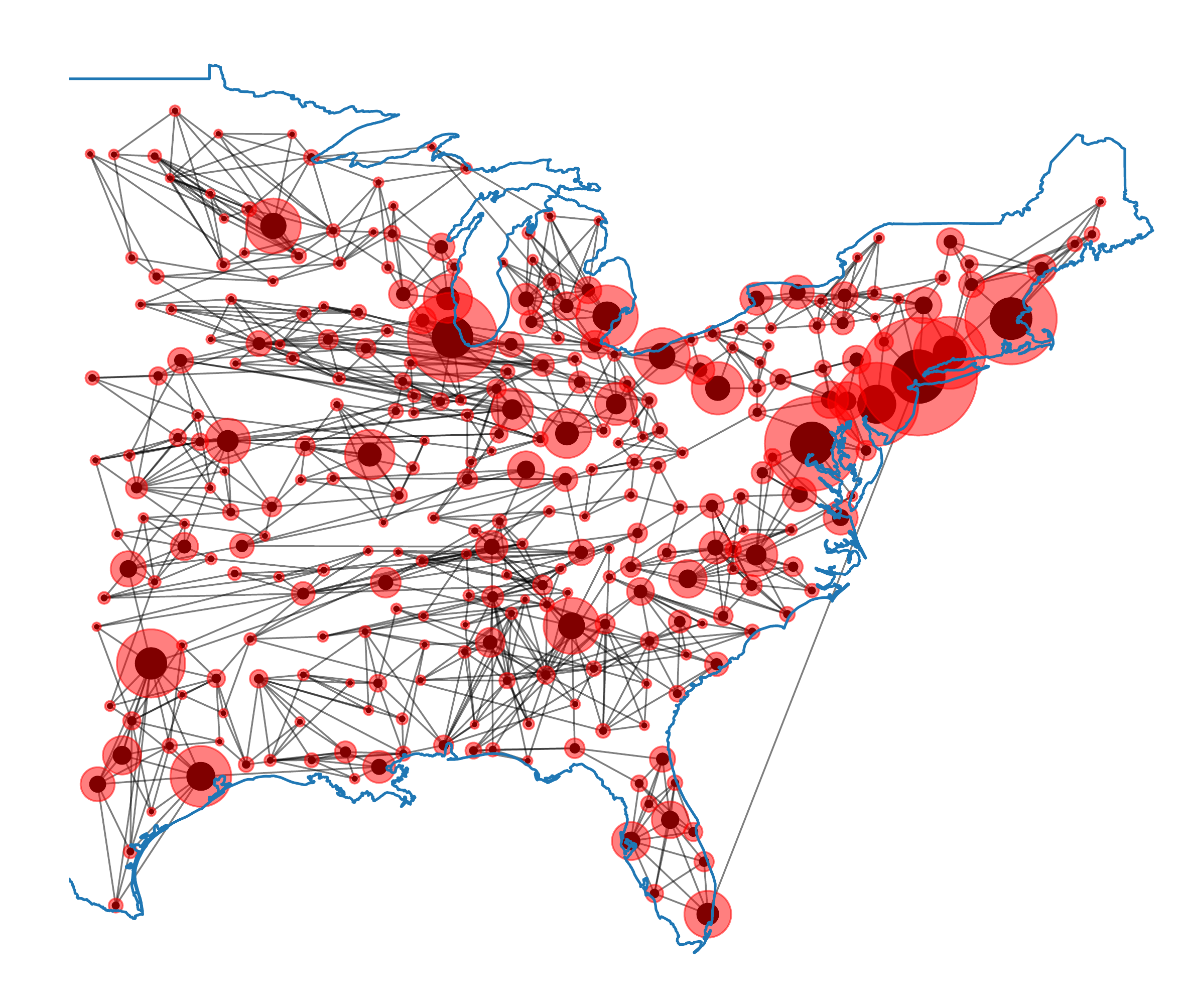}
	\caption{\label{fig:network} Linguistic proximity network on the set of population nodes using survey data. Each node is connected to the four nodes to which is most similar linguistically.    }
\end{figure}

To further explore the relationship between linguistic and geographical proximity in the USA, in Figure \ref{fig:network} we have constructed a network on our set of $N=300$ nodes in which every node connects to its four closest linguistic neighbours. 
Note that most connections are short range compared to the system size; population centres are typically most linguistically similar to others within a few hundred kilometres with the striking exception of Miami, Florida (which will come as no surprise to linguists who are well aware of  the strong influence on Florida of migration from North Eastern US cities \cite{sam12,she91,cens11}). The social network through which linguistic forms spread may therefore be viewed as quasi two-dimensional, provided we take a sufficiently coarse grained view of the system. This has geometrical implications for the conformity driven evolution of language; if the social network over which language evolves is two dimensional, then linguistic boundaries may be viewed as lines and by analogy with conformity driven physical systems, we would expect these to feel surface tension \cite{bra94,bur17,bur18}. 
\begin{figure}
	\includegraphics[width=\columnwidth]{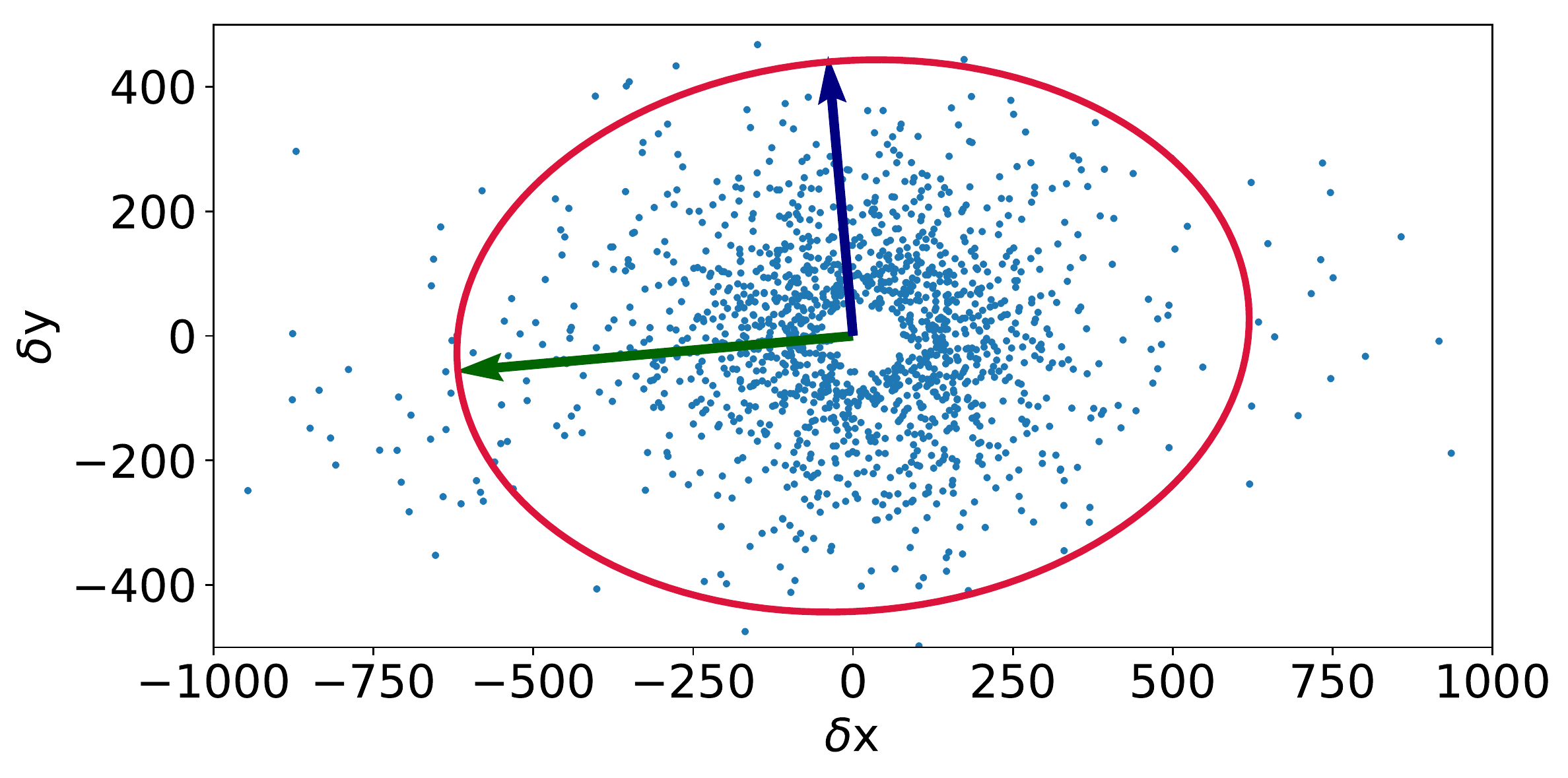}
	\caption{\label{fig:coords} Set of relative geographical coordinates $\bv{r}_i-\bv{r}_j$ for all pairs of nodes $(i,j)$ where $j$ is one of the four linguistically closest nodes to $i$. Arrows show the eigenvectors of the gyration tensor (\ref{eqn:inertia}) for the set of displacements. Red ellipse encloses $95\%$ of data and has major and minor axes proportional to corresponding eigenvalues. The length ratio of the major to the minor axis 1.40.    }
\end{figure}
We also observe from Figure \ref{fig:network} that the distribution of connections is not isotropic: a disproportionate number of edges appear to run closer to the east-west direction than to north-south. To explore this effect, in Figure \ref{fig:coords} we have plotted the relative geographical coordinates of all pairs of nodes connected in Figure \ref{fig:network}. Also shown in Figure \ref{fig:coords} are the eigenvectors of the gyration tensor
\begin{equation}
\label{eqn:inertia}
T = \begin{bmatrix}
\la \Delta x ^2 \ra & \la \Delta x \Delta y \ra \\
\la \Delta x \Delta y \ra & \la \Delta y^2 \ra
\end{bmatrix}
\end{equation} 
where $\Delta x, \Delta y$ are the relative displacements in the east-west and north-south directions respectively. The eigenvectors of $T$ are closely aligned with the north-south and east-west directions, and the ratio of the corresponding eigenvalues is 1.40. There is therefore strong anisotropy in the geographical distribution of linguistic near-neighbours. In other words, the drop in cultural and linguistic affinity between population settlements appears to decline more slowly with east-west displacements than north-south. It is possible that this anisotropy is a historical artefact of the west-moving colonisation of the continent \cite{kur28, wol16}, leading to disproportionately strong east-west cultural identification, or it may be due to the existence of more extensive east-west oriented physical routes of communication (e.g. roads, air flights), or even a combination of both these things. We explore this possibility below.

\section{The model}

Our aim is to test the extent to which spatial linguistic patterns can be explained by minimal statistical physics models of conformity--driven evolution. This follows on from the language models defined in \cite{bur17, bur18}, in which the spatial evolution of linguistic memory was governed by a modified time-dependent Ginzburg-Landau equation, the original purpose of which was to provide a coarse grained (continuous) description of coarsening in physical systems \cite{bra94}. Since we are interested in the behaviour of a network, in the current paper we revert to the simplest possible model whose steady states are a discrete analogue of those generated by the continuous space memory models defined in \cite{bur17,bur18}. Our model is a generalisation of the $q$-state Potts model \cite{pot52}, described below. Our aim is not to capture the full stochastic evolution of the frequency vectors $\bv{f}_i^Q$ (to be considered in further work), but to model the discrete patterns which emerge from the survey data after clustering. We note that there exists a large body of work exploring the processes which drive linguistic change \cite{lab01,tru74,cha98,wol03,blo33,cro00}, and factors such as gender, networks (social and physical), social status and cultural identities all play important roles. 

\begin{figure}
	\includegraphics[width=\columnwidth]{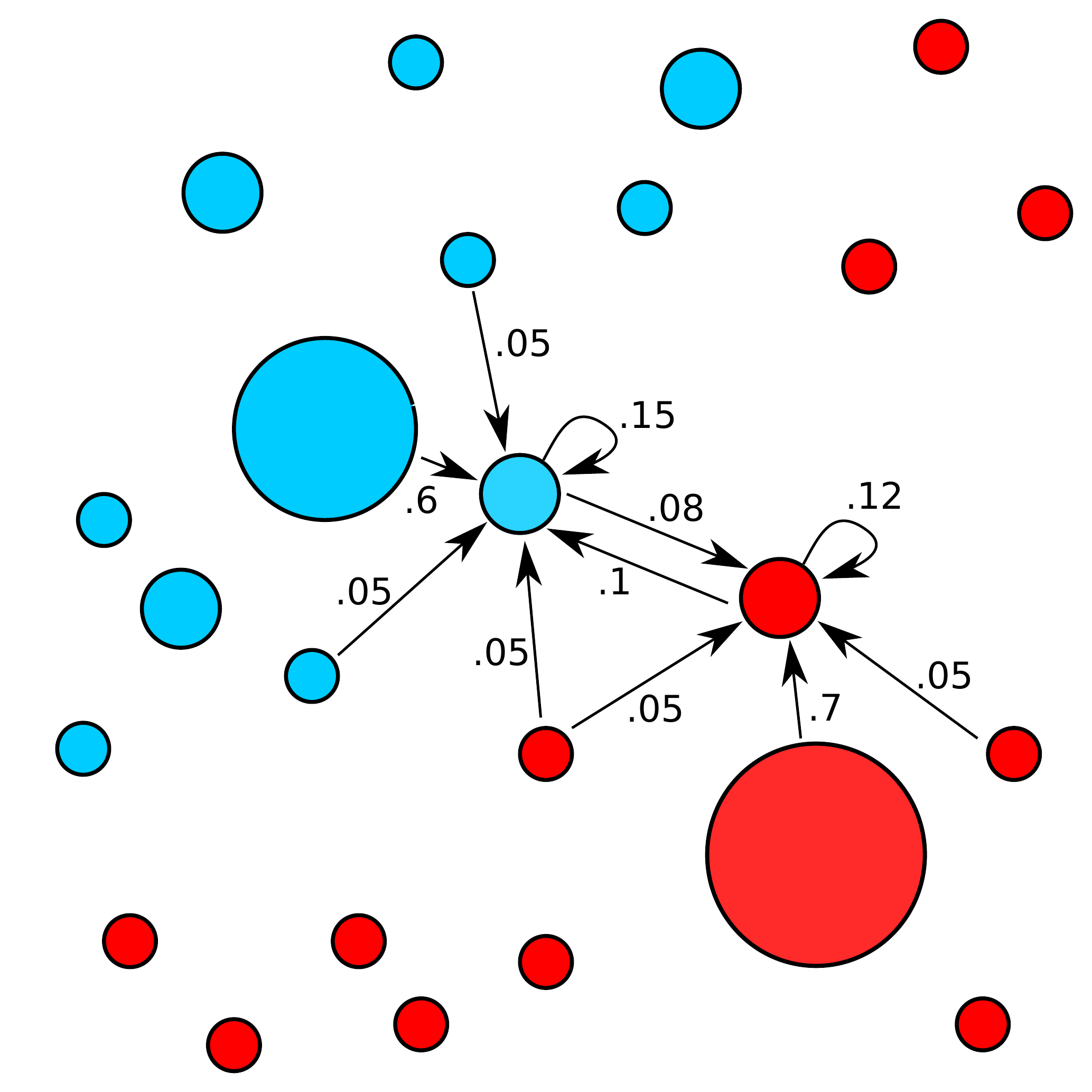}
	\caption{\label{fig:potts} Schematic diagram of the influences on two different nodes. Influence values are for illustrative purposes: nodes feel the greatest influence from others which are large and close.      }
\end{figure}

Our model is defined as follows. At time $t$, each node $i$ of the network is in a discrete state $s_i(t) \in \{1,2,\ldots,q\}$. We relate this discrete state to our continuous observational data by viewing $s_i$ as the label of the linguistic cluster assigned to $\bv{f}_i^Q$ by our clustering method of choice. We assume that nodes evolve so as to maximize conformity within their linguistic neighbourhood. We define this neighbourhood using a discrete version of the interaction kernel defined in \cite{bur17}. Letting $\bv{r}_i$ be the location of node $i$, we first define a raw kernel $\phi(\bv{r}_i-\bv{r}_j)$ giving the influence of node $i$ on node $j$ in the absence of variations in the populations of nodes. We then define the normalized population weighted influence of node $j$ on node $i$ to be
\begin{equation}
J_{ij} = \frac{ \phi(\bv{r}_i-\bv{r}_j) P_j}{\sum_k \phi(\bv{r}_i-\bv{r}_k) P_k} .
\label{eqn:J}
\end{equation}  
According to this definition, the influence of a node scales in proportion to its population $P_j$: if two nodes are equidistant from a speaker, she is twice as likely to converse with another speaker from a node with twice the population. We note that the idea of modelling linguistic change using population-based measures of influence 
%(different to \ref{eqn:J}) 
was introduced by the sociolinguist Peter Tudgill in his gravity model \cite{cha98,tru74}, which predicts how changes jump from one settlement to another. Here we take a different approach by defining dynamics which seek to minimize a global non-conformity function, analogous to the Hamiltonian of the $q$-state Potts model. We first define the indicator function that two nodes belong to different clusters
\begin{equation}
\Delta_{ij}(t) = \begin{cases}
0 & \text{ if } s_i(t) = s_j(t) \\
1 & \text{ otherwize}.
\end{cases}
\end{equation}
We call this the indicator of non-conformity between nodes. The total non-conformity of the entire system is then
\begin{equation}
H(t) = \sum_{i,j} J_{ij} \Delta_{ij}(t).
\end{equation}
We note that the influence numbers $J_{ij}$ are not symmetric in their indices because larger nodes will exert greater linguistic influence on their smaller neighbours than these neighbours exert in return. 

Conformity driven dynamics are then implemented using the Metropolis algorithm \cite{pot52}: we randomly select a node $i$, and then propose a new state, $s'$, selected uniformly at random from the set of alternatives to the current state $s_i(t)$. We let $\Delta H$ be the change in total non-conformity that would result from the proposed change, and then accept this change with probability
\begin{equation}
\PP(s_i(t) \rar s') = \begin{cases}
1 & \text{ if } \Delta H <0 \\
e^{-\beta \Delta H} & \text{ if } \Delta H \geq 0,
\end{cases}
\end{equation}
where $\beta$, the classical inverse temperature, controls the level of noise in the dynamics. In the zero noise (zero temperature) limit $\beta \rar \infty$, only those changes which increase conformity are allowed. For simplicity we consider the zero temperature limit from here on. Although these dynamics are minimal and coarse grained, they capture an important aspect of spatially distributed social ordering phenomena: namely that even if the individual behaviour of agents tends to lead toward social conformity, the system as a whole may exhibit regionalism because the stochastic process of conforming takes place at small spatial scales. The system can therefore become ``stuck'' in a suboptimal global state, because no single change of individual state can increase conformity.

%After each Metropolis step we increment time by $\dt = 1/N$ so that the average update rate for each node is unity. 

In order to allow for anisotropy in our interaction kernel, we define the anisotropic distance $d_{ij}$ between nodes as 
\begin{equation}
d_{ij}(\bv{r}_i-\bv{r}_j) = \sqrt{\frac{(x_i-x_j)^2}{A^2} + \frac{(y_i-y_j)^2}{1^2}} 
\end{equation}
where $A$ measures the extent to which the east-west components of geographical displacements are effectively shrunk by enhanced connectivity (cultural or physical). Setting $A>1$ is equivalent to squashing the system in the east-west direction. Given the anisotropic distance, we define our raw interaction kernel $\phi$ to be a truncated Cauchy distribution 
\begin{equation}
\phi(\bv{r}_i-\bv{r}_j) = \begin{cases}
\left(1 + \left(\frac{d_{ij}}{\gamma}\right)^2 \right)^{-1} & \text{ if } d_{ij} < R \\
0 & \text{ otherwise}
\end{cases}
\label{eqn:cauchy}
\end{equation} 
where $\Delta r_{ij} = |\bv{r}_i-\bv{r}_j|$ is the distance between nodes $i$ and $j$. We refer to $\gamma$ as the \textit{interaction range} and $R$ as the \textit{cut-off}.  Our choice of $\phi$ is in part guided by experimental data which suggest that human displacements collectively follow a truncated power-law, but are individually highly repetitive and predictable \cite{gon08,son10}, with considerable heterogeneity within the population. We also note that the choice of a long range algebraic kernel as opposed to a short range exponentially decaying kernel is more consistent with our linguistic proximity network (Figure \ref{fig:network}), which contains some links which stretch hundreds of kilometres. The large distance cut-off may also be justified on purely theoretical grounds since without it, inverse square-law interactions preclude the possibility of stable domain walls in phase ordering models \cite{bla13}.

\section{Results}

\begin{figure}
	\includegraphics[width=\columnwidth]{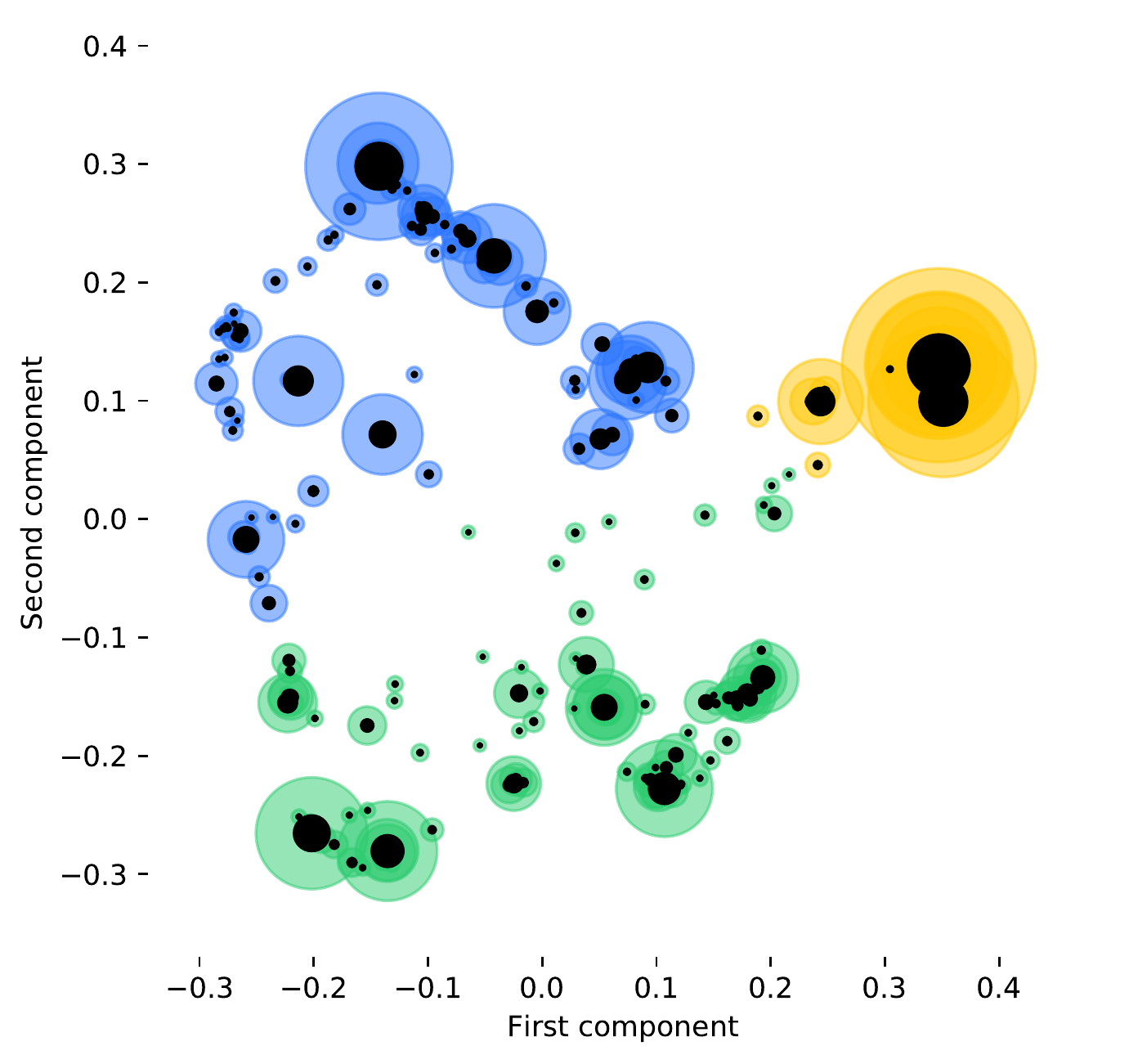}
	\caption{\label{fig:pca_model} Principle components analysis of aggregated Potts results with $R=500$km and $\gamma=120$km. Figure shows a typical output, variations occur between simulation runs.     }
\end{figure}

\begin{figure}
	\includegraphics[width=\columnwidth]{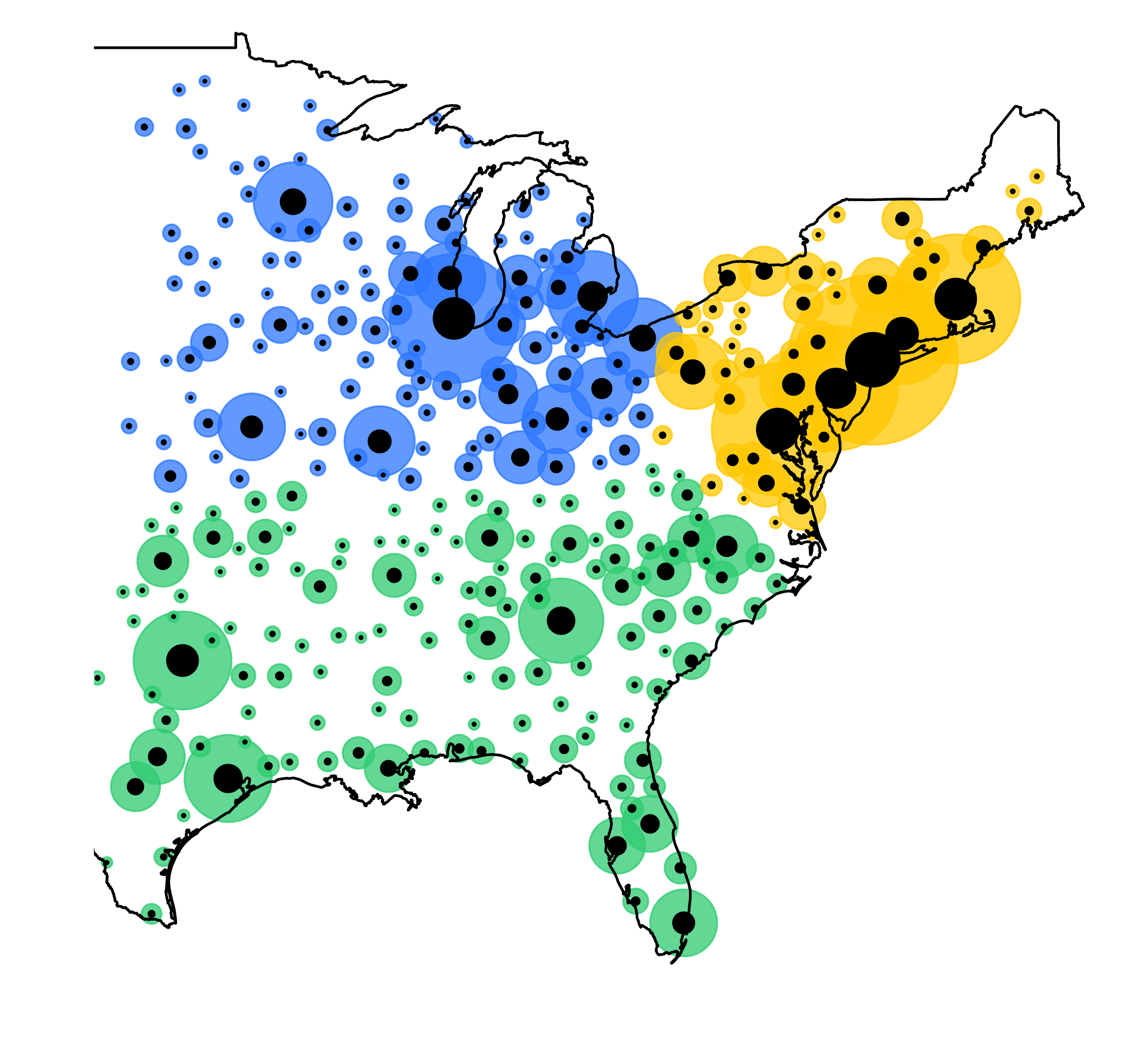}
	\caption{\label{fig:clus_model} Cluster analysis of aggregated Potts results with $R=500$km and $\gamma=120$km.   }
\end{figure}

%We initially explore the behaviour of the model with $R=400$km, consistent with \cite{gon08}. 
Applying the silhouette method \cite{rou87} to each of the survey questions, we find that the optimal number of clusters in all cases is either $q=2$ or $q=3$. We therefore explore the behaviour of the two- and three-state versions our model over a large number of simulation runs. Since we have no information regarding the early linguistic state of our population nodes, we initialise the system with each node in a random state $s_i(0) \in \{1, \ldots , q\}$. We note that the number of possible initial conditions is very large indeed $> 10^{90}$ -- significantly larger than the number of atoms in the universe. This approach may be justified on the basis that although these initial conditions almost certainly do not reflect reality, the ordering process causes very large numbers of different early states to converge to a much smaller subset of equilibrium configurations. In order to estimate the number of these \textit{attractors} of the dynamics we require a simple method for measuring the similarity of two maps. Two clusterings of nodes are equivalent if they can be transformed into one another by permuting cluster labels. Therefore, in order to compare maps we must find the permutation of labels which maximizes the number of nodes which have the same label in both maps. This can be achieved using the Hungarian algorithm \cite{kuh55}, and results in a similarity score, $S(m_1,m_2)$, giving the fraction of nodes which belong to the same cluster in map $m_1$ and $m_2$. We may then generate a set, $\mathcal{D}$, of distinctive maps by repeatedly generating a new map, $m$, and adding it to $\mathcal{D}$ only if $S(m,m_i) < 90\%$ for all $m_i \in \mathcal{D}$.
For small interaction ranges $|\mathcal{D}|$ can be very large but for the range of parameter values used in this work $|\mathcal{D}|$ is only a few hundred. Since all initial conditions are attracted toward this set, we argue that that the current linguistic state of the system may still be significantly predictable even though we lack historical data; it doesn't matter where you start, because in the end you will end up near to $\mathcal{D}$. 

To make our predictions we generate $750$ Potts maps for each $q \in \{2,3\}$ using $3 \times 10^4$ metropolis updates (sufficient to reach a near-equilibrium state). Letting $s_i^k$ be the final state of the node $i$ in map $k$ we define linguistic state of node $i$ to be the vector of states for all maps
\begin{equation}
\bv{s}_i = (s_i^1,s_i^2, \ldots ).
\end{equation}
The first two principle components of these vectors, obtained from simulations are shown in Figure \ref{fig:pca_model}. In order to compare the model to our survey data we use $K$-means to assign each node to one of three clusters based purely on its linguistic state, and then plot these clusters geographically in Figure \ref{fig:clus_model}. Visual inspection reveals a close match to the survey map in Figure \ref{fig:all_data}, confirmed by a similarity score of $S(\text{model},\text{data})=90.67\%$.

To test the robustness of this result to variations in interaction kernel parameters we our repeat simulations and analysis for $R\in \{400,500,600\}$ and $\gamma \in \{90,100,\ldots,180\}$. The results in the isotropic case ($A=1$) are shown in Table \ref{tbl:A1}, where we see that a similarity score over $80\%$ is achieved in 13 out of 30 cases.  These high scoring maps all exhibit approximately the same pattern illustrated in Figure \ref{fig:clus_model}. In cases when this pattern fails to emerge the system selects one of a number of alternative states, and the score drops considerably.  When interaction range is excessively low, nodes become increasingly independent, and linguistic clusters either do not form, or are very small.

To investigate the effects of interaction anisotropy, we stretch the interaction kernel in the east-west direction by setting $A=1.15$ and repeat our previous analysis (see Table \ref{tbl:A115}). In this case we obtain a robust match between model and survey maps for all parameter values. Our analysis of the linguistic proximity network (Figures \ref{fig:network} and \ref{fig:coords}) suggested a greater east-west affinity, and our results indicate that such an affinity causes the observed linguistic zones  to appear with high probability in our model. That fact that this pattern is very robust with respect to the choice of interaction parameters suggests that the observed patterns may be explained by the distribution of people, and by system shape, rather then the fine detail of their interactions. In fact, it is likely that regional differences in connectivity and population density would make the interaction range vary by location. Comparing maps generated for individual survey questions to the set of simulated maps, we find the best-match between survey and simulated maps, averaged over all survey questions is $\approx 80\%$. If we stretch our kernel in the oppose direction by setting $A=0.96$, we find no strong matches to our survey maps (Table \ref{tbl:A096}).

%proper distances
\begin{table}
	\begin{tabular}{|l|l|l|l|l|l|l|l|l|l|l|}
		\hline
		\backslashbox{$R$}{$\gamma$}  	& \textbf{90}                 & \textbf{100}                & \textbf{110}                & \textbf{120}                & \textbf{130}                & \textbf{140}                & \textbf{150}                & \textbf{160}                & \textbf{170}                & \textbf{180}                \\ \hline
		\textbf{400} & {\color{blue} 0.87} &  0.53 &  0.58 &  0.53 & {\color{blue} 0.85} & {\color{blue} 0.84} & 0.55 &  0.58 &{\color{blue}0.89} & {\color{blue} 0.85} \\ \hline
		\textbf{500} & {\color{blue} 0.88} & {\color{blue} 0.9}  & 0.75 & {\color{blue} 0.9}  & {\color{blue} 0.87} &  0.63 & {\color{blue} 0.89} & 0.75 & 0.66& {\color{blue} 0.89} \\ \hline
		\textbf{600} & 0.75 & {\color{blue} 0.9}  & {\color{blue} 0.87} & 0.75 &  0.75 & 0.67 &0.67 &  0.67 & 0.67 & 0.67 \\ \hline
	\end{tabular}
	\caption{\label{tbl:A1} Percentage similarity scores between model and data for isotropic interaction kernel ($A=1$). Scores above $80\%$ are coloured blue, otherwise black.}
\end{table}

\begin{table}
	%good stretch
	\begin{tabular}{|l|l|l|l|l|l|l|l|l|l|l|}
		\hline
		\backslashbox{$R$}{$\gamma$}  	& \textbf{90}                 & \textbf{100}                & \textbf{110}                & \textbf{120}                & \textbf{130}                & \textbf{140}                & \textbf{150}                & \textbf{160}                & \textbf{170}                & \textbf{180}                \\ \hline
		\textbf{400} & {\color{blue} 0.82} & {\color{blue} 0.9}  & {\color{blue} 0.89} & {\color{blue} 0.89} & {\color{blue} 0.89} & {\color{blue} 0.89} & {\color{blue} 0.89} & {\color{blue} 0.87} & {\color{blue} 0.89} & {\color{blue} 0.89} \\ \hline
		\textbf{500} & {\color{blue} 0.89} & {\color{blue} 0.85} & {\color{blue} 0.89} & {\color{blue} 0.85} & {\color{blue} 0.89} & {\color{blue} 0.89} & {\color{blue} 0.85} & {\color{blue} 0.85} & {\color{blue} 0.85} & {\color{blue} 0.85} \\ \hline
		\textbf{600} & {\color{blue} 0.88} & {\color{blue} 0.85} & {\color{blue} 0.85} & {\color{blue} 0.84} & {\color{blue} 0.84} & {\color{blue} 0.83} & {\color{blue} 0.84} & {\color{blue} 0.84} & {\color{blue} 0.84} & {\color{blue} 0.82} \\ \hline
	\end{tabular}
	\caption{\label{tbl:A115} Percentage similarity scores between model and data for anisotropic interaction kernel with $A=1.15$ (east-west displacements shrunk). Scores above $80\%$ are coloured blue, otherwise black.}
\end{table}

\begin{table}[h!]
	%badly stretched
	\begin{tabular}{|l|l|l|l|l|l|l|l|l|l|l|}
	\hline
	\backslashbox{$R$}{$\gamma$} 	& \textbf{90}                 & \textbf{100}                & \textbf{110}                & \textbf{120}                & \textbf{130}                & \textbf{140}                & \textbf{150}                & \textbf{160}                & \textbf{170}                & \textbf{180}                \\ \hline
	\textbf{400} & {\color{black} 0.63} & {\color{black} 0.63} & {\color{black} 0.6}  & {\color{black} 0.6}  & {\color{black} 0.6}  & {\color{black} 0.58} & {\color{black} 0.63} & {\color{black} 0.6}  & {\color{black} 0.63} & {\color{black} 0.63} \\ \hline
	\textbf{500} & {\color{black} 0.61} & {\color{black} 0.61} & {\color{black} 0.61} & {\color{black} 0.6}  & {\color{black} 0.6}  & {\color{black} 0.6}  & {\color{black} 0.63} & {\color{black} 0.64} & {\color{black} 0.66} & {\color{black} 0.66} \\ \hline
	\textbf{600} & {\color{black} 0.62} & {\color{black} 0.62} & {\color{black} 0.63} & {\color{black} 0.62} & {\color{black} 0.66} & {\color{black} 0.63} & {\color{black} 0.64} & {\color{black} 0.66} & {\color{black} 0.66} & {\color{black} 0.63} \\ \hline
\end{tabular}
	\caption{\label{tbl:A096} Percentage similarity scores between model and data for anisotropic interaction kernel with $A=0.96$ (east-west displacements increased). Scores above $80\%$ are coloured blue, otherwise black.}	
\end{table}

\begin{figure}[h!]
	\includegraphics[width=\columnwidth]{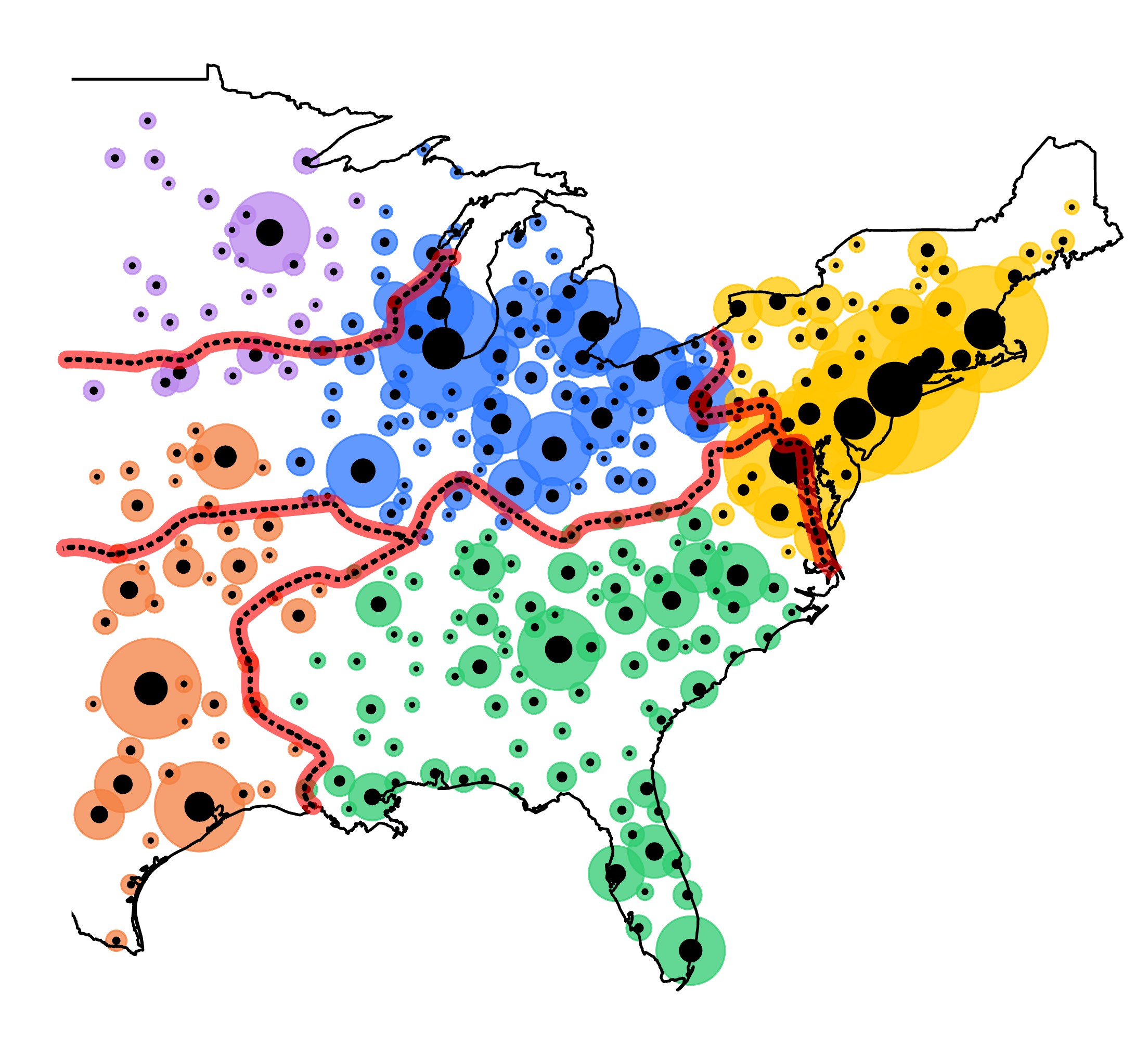}
	\caption{\label{fig:twit} $K$-means clustering of aggregated Potts result into $K=5$ clusters. Red (dashed) lines give boundaries of 5 clusters found from analysis of Twitter data by Huang et. al. \cite{gri16}.}.
\end{figure}

As an alternative to dedicated linguistic surveys, social media platforms such as Twitter may be used to generate very large datasets of geo-tagged text which may be analysed to discover geographical variations in language use \cite{gri16}. A recent analysis of 924 million tweets  generated by 6.6 million users in the USA over one year used hierarchical clustering to divide the country into distinctive linguistic zones. For example, the five most distinctive clusters are shown in Figure \ref{fig:twit}, along with the aggregated result of our Potts model, divided into five clusters using $K$-means. We note the close match, and also that clusters appear to have densely populated areas at their heart with boundaries lying in less densely populated areas. These features were predicted by the memory based surface tension models \cite{bur17,bur18} upon which the current paper builds.

%\begin{figure}
%	\includegraphics[width=\columnwidth]{pca_voronoi.pdf}
%	\caption{\label{fig:pca_vor} Principle components analysis of aggregated Voronoi clusters. Colours indicate K-means clusters.  }
%\end{figure}

%\begin{figure}
%	\includegraphics[width=\columnwidth]{geoclusters_voronoi.pdf}
%	\caption{\label{fig:vor} Cluster analysis of aggregated Voronoi results.   }
%\end{figure}

\section{Voronoi null model}

\begin{figure}
	\includegraphics[width=\columnwidth]{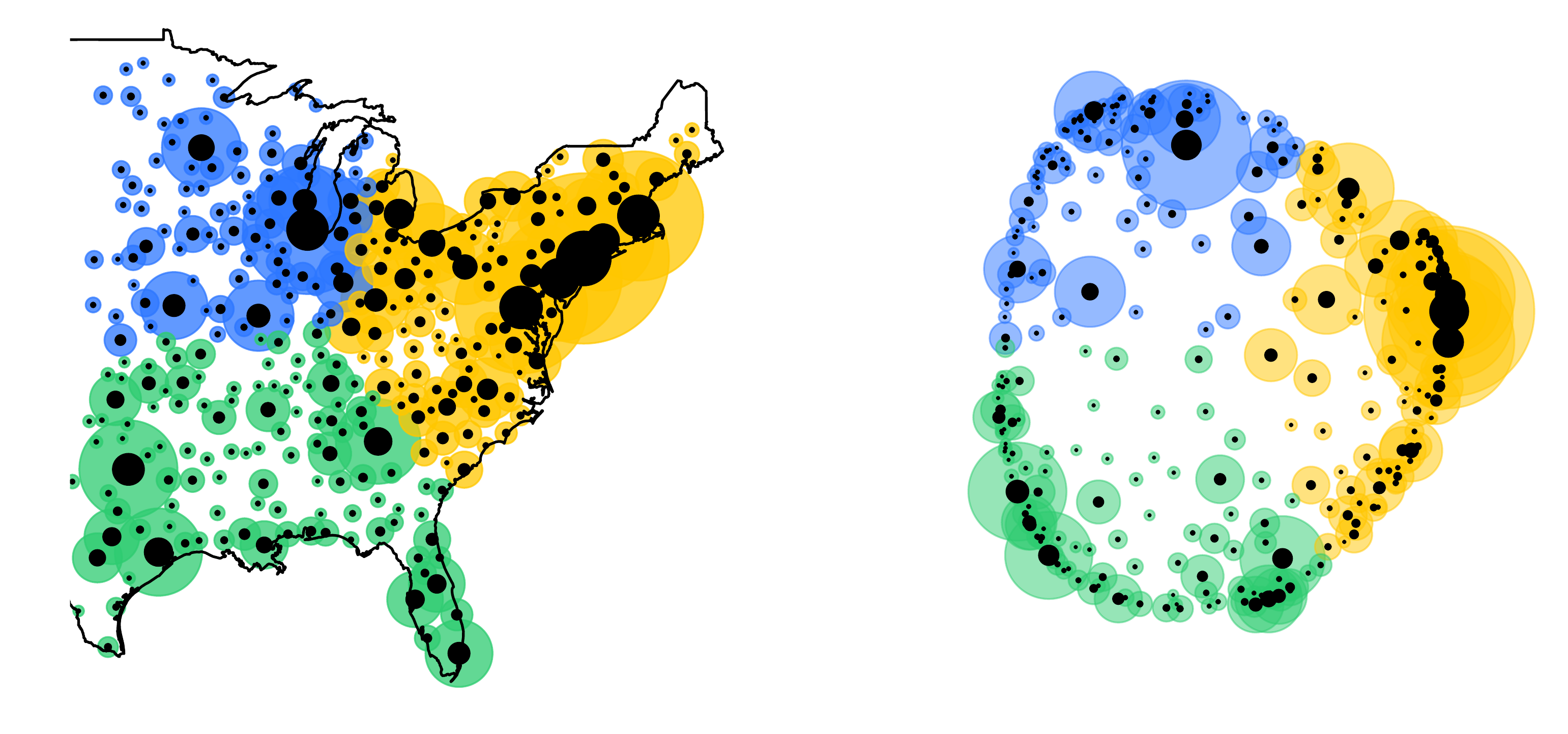}
	\caption{\label{fig:vor} Left: Cluster analysis of aggregated Voronoi results.  Right: Principle components analysis of aggregated Voronoi clusters. Colours indicate $K$-means clusters. This map produces a 70\% match to survey cluster map.   }
\end{figure}

In defining our model we took account of, and made assumptions regarding, the effect of social conformity and the relative sizes of population centres. In order to assess the extent to which these factors are necessary to explain the observed geographical linguistic patterns, or whether the patterns are the result of a more trivial process, we now define a simple null model of linguistic clustering. The only assumption of this simpler model is that nearby nodes should be linguistically similar. A simple way to achieve this is to select $q$ nodes at random, assign each a different label, and then assign all other node labels according to which of the original $q$ nodes is closest. In this way we generate a discrete version of the Voronoi tessellation \cite{chi13}. Having generated a large number of tessellations, each representing the null-map for a single fictitious survey question, we repeat our principle components and cluster analysis. The principle components plot (Figure \ref{fig:vor}) takes the form of a circle, lacking any recognisable clusters in linguistic space. This high degree of symmetry reflects that fact that the null model treats all nodes as equivalent in size and status, and ensures only that geographically nearby nodes are close linguistically. 
%In this sense the null model implements a primitive form of social conformity, but without accounting for population size or specifying  any form of dynamics.  
Despite the lack of clearly identifiable clusters, we may still apply $K$-means clustering which seeks to minimize within cluster variation \cite{has09}. The result is a symmetrical division of points into approximately equal sized clusters, shown in Figure \ref{fig:vor}. 
%Since linguistic proximity is a simple proxy for geographical proximity within the null model, we expect K-means to produce similar results when applied to linguistic states or geographical positions. 

There is similarity between shapes of the main linguistic zones in our Voronoi null model the survey map (Figures \ref{fig:vor} and \ref{fig:all_data}). This similarity is not surprising if one accepts that people nearby tend to use similar linguistic terms. This assumption was encoded in the null model. Our Potts model shows that deviations from this null picture, due to social conformity and uneven population distribution, produce clustering in linguistic space, present in the survey data and Potts model (Figures \ref{fig:pca} and \ref{fig:pca_model}) but not in the null model. Beyond linguistic clustering, the Potts model also produces a closer match to the survey map. We therefore suggest that these two effects are a necessary ingredient to understand the distribution of language features. A discussion of their effects in the continuous space setting is given in \cite{bur17,bur18}.

\section{Discussion}

We have examined the spatial distribution of linguistic features in eastern USA, and compared these distributions to a generalized Potts model defined on the network of population centres, taking account of long-range interactions, social conformity, population sizes and interaction anisotropy.  The steady states of this model are discrete analogues of those generated by the continuous space memory models defined in \cite{bur17,bur18}, where the dialect patterns of England and Italy were modelled.  

In the case of isotropic interactions, our Potts model predicted shapes of linguistic zones in close ($ \approx 90\% $) agreement with survey data \cite{vau17} for 13 out of 30 sets of interaction parameters tested. Analysis of the linguistic proximity network inferred from survey data suggested that linguistic affinity was not in fact isotropic, and introducing such anisotropy into our model produced a close agreement to the survey maps for \textit{all} parameter values. 

%In this way we have provided further evidence that the shape of a country and the distribution of its people have powerful and predictable effects on the large scale spatial distribution of linguistic features \cite{bur17,bur18}. 

It is interesting to  note a possible connection to crossing probabilities in percolation \cite{bar09}. In low temperature two dimensional magnetic systems it is common to see \textit{stripe states} formed by magnetic domain walls crossing the system. The appearance and direction of these crossings depends strongly on the aspect ratio of the system: they become increasingly common as the aspect ratio is increased, and in high ratio system they typically run across the short axis. By shrinking east-west displacements we were effectively changing the aspect ratio of our system, making an east west domain wall substantially more likely.

%Our work contributes a partial answer to the question of whether the north-south linguistic divide in the United States is a historical artefact of westerly colonisation \cite{kur28}, or simply a consequence of population distribution and geography. 

Our model indicates that without anisotropy, the current population distribution could have generated a linguistic north-south split, but this distribution is only one of a number of possibilities. By including anisotropy we find that the split is almost inevitable. Our work therefore takes a step towards answering the question of whether the observed north-south linguistic divide in the USA is merely a consequence of population distribution and geography. Our results support the idea that enhanced east-west linguistic transmission has occurred in the USA. Enhanced east-west transmission could have arisen from the historic westerly colonization (migration) of people. Alternatively the existence of better east-west transportation links, either historically or in the modern setting, could provide an explanation, and in future work we might analyze historic and modern transportation networks to test this possible explanation. However we note that the two possible explanations (migration and the quality or transportation links) are often inter-dependent.

%Future work may be able to uncouple the extent to which this stronger east-west transmission is due to historic westerly migrations and/or due to the pattern of road or other communication networks in the historic or modern USA. 

%It is possible that the distribution of people is itself a consequence of the north-south cultural divide, or that deeper analysis of more linguistic data sets will demonstrate that the geographical distribution of language requires west moving colonisation for its explanation. We aim to address this in future work.    

\acknowledgments
\noindent
J.B is grateful for the support of a Royal Society APEX award 2018-2020. The authors are grateful to Marius J\o hndal for curating and supplying the survey data. 

\subsubsection*{Author contributions}
The individual contributions of the authors were: J.B. devised the study and model, wrote simulations, performed data analysis and wrote manuscript. B.V devised original survey, provided data and linguistic expertise, and edited manuscript. M.G. processed and analysed data, ran simulations, and generated figures and tables. Y.G. performed exploratory simulations and data analysis, directed by J.B.

\bibliography{swe_potts_refs}

\end{document}